\documentstyle[12pt]{article}

\begin{document}
\pagestyle{empty}

%********************  User defined commands  ***************************
\newcommand{\eq}[1]{eq.~(\ref{#1})}
\newcommand{\eqs}[2]{eqs.(\ref{#1}, \ref{#2})}
\newcommand{\eqss}[3]{eqs.(\ref{#1}, \ref{#2}, ref{#3})}
\newcommand{\ur}[1]{~(\ref{#1})}
\newcommand{\urs}[2]{~(\ref{#1},~\ref{#2})}
\newcommand{\urss}[3]{~(\ref{#1},~\ref{#2},~\ref{#3})}
\newcommand{\Eq}[1]{Eq.~(\ref{#1})}
\newcommand{\Eqs}[2]{Eqs.(\ref{#1},\ref{#2})}
\newcommand{\fig}[1]{Fig.~\ref{#1}}
\newcommand{\figs}[2]{Figs.\ref{#1},\ref{#2}}
\newcommand{\figss}[3]{Figs.\ref{#1},\ref{#2},\ref{#3}}
\newcommand{\beq}{\begin{equation}}
\newcommand{\eeq}{\end{equation}}
\newcommand{\e}{\varepsilon}
\newcommand{\ee}{\epsilon}
\newcommand{\bI}{\bar I}
\newcommand{\eoe}{(\bar{\eta}O\eta)}
\newcommand{\ui}{U_{int}(R,\rho_{1,2},O_{1,2})}
\newcommand{\thb}[3]{\bar{\eta}^{#1}_{#2 #3}}
\newcommand{\tha}[3]{\eta^{#1}_{#2 #3}}
\newcommand{\la}[1]{\label{#1}}
\newcommand{\YM}{Yang--Mills~}
\newcommand{\N}{$N_{CS}\;$}
\newcommand{\Nz}{$N_{CS}=0\;$}
\newcommand{\No}{$N_{CS}=1\;$}
\newcommand{\x}{{\bf x}}
\newcommand{\r}{{\bf r}}
\newcommand{\IIs}{$I$'s and ${\bar I}$'s }
\newcommand{\II}{$I{\bar I}$ }
%********************************************

\vskip 1true cm

\begin{center} \Large\bf
Foundations of the Constituent Quark Model
\end{center}

\vskip 1true cm

\begin{center} \large Dmitri Diakonov \end{center}

\vskip .7true cm

\begin{center}
\small\it Petersburg Nuclear Physics Institute, Gatchina, St.Petersburg
188350, Russia \end{center}

\vskip 2true cm

\hskip 1.5true cm {\bf Abstract}

`Constituent quarks' means massive quarks, in contrast to the
nearly massless $u,\;d,\;s$ quarks of the QCD lagrangian. The dynamical
or constituent masses appear owing to the spontaneous chiral
symmetry breaking in QCD. A realistic mechanism of the chiral symmetry
breaking is provided by instantons. Therefore, I review the present
status of the QCD instanton vacuum, emphasizing the mechanism of chiral
symmetry breaking and of the generation of the constituent quark mass.
I end up with the Nambu--Jona-Lasinio-type effective theory to which
QCD is reduced at low momenta.

\vskip .7true cm
\hskip 1.5true cm {\bf Keywords}

Constituent quarks, chiral symmetry breaking, instantons, effective chiral
lagrangian, Nambu--Jona-Lasinio model

\vskip 1true cm

\section{Introduction}

Quantum Chromodynamics (QCD) is the theory of strong interactions. From
a theorist's point of view strong interactions means that all
dimensionless quantitites of the theory are, generally speaking, of the
order of unity. Meanwhile, there are at least two fundamental facts about
strong interactions showing that actually it is not exactly so: there
are still certain small parameters hidden. One is that nuclei can be, to
a good accuracy, viewed as made of weakly bound nucleons --- in a true
`strong interactions' case they would rather look like a quark soup. The
second puzzle is that the nucleon itself can be, to a good accuracy,
viewed as built of three constituent quarks --- in a true `strong
interactions' case it would rather look like a pack of an indefinite
number of quarks, antiquarks and gluons.

The second puzzle (and probably the first, too) is qualitatively
explained by the spontaneous breaking of the (approximate) chiral
symmetry of QCD. The bare ot current masses of the light quarks
($m_u\simeq 4\;MeV, \;\;m_d\simeq 7\;MeV, \;\;m_s\simeq 150\;MeV$)
are small as compared to the typical scale of strong interactions,
say, the $\rho$-meson mass, $770\;MeV$, and can be put to zero.
In that limit (called the chiral limit) the QCD lagrangian possesses
an 18-parameter symmetry, called chiral symmetry. It is the symmetry
under independent $U(3)\times U(3)$ rotations of the left- and
right-handed components of the $u,\;d,\;s$ fields. Equivalently,
it is the symmetry under $U(3)_{Vector}\times U(3)_{Axial}$ rotations.
Were this symmetry exact, we would observe degeneracy between states
with opposite parity but otherwise the same quantum numbers. For example,
the vector $\rho$-meson would be degenerate with the axial $a_1$-meson;
the nucleon $(\frac{1}{2}^+,\;940)$ would be degenerate with the
resonance $(\frac{1}{2}^-,\;1535)$, etc. Since it is not the case, we
conclude that the $U(3)_A$ symmetry is broken {\em spontaneously}.
The order parameter for symmetry breaking is the quark or {\em chiral
condensate},

\beq
\langle \bar\psi\psi\rangle\simeq -(250\;MeV)^3, \;\;\;\;\psi=u,d,s.
\la{chcond}\eeq
Its $U(3)_A$ rotations produce nine massless Goldstone bosons --
the pseudoscalar nonet.

It should be noted that the above quantity is well defined only for
massless quarks, otherwise it is somewhat ambigious. By definition,
this is the quark Green function taken at one point; in momentum space
it is a closed quark loop. Had the quark propagator only the "slash"
term, the trace over the spinor indices understood in this loop would give
an identical zero. Therefore, chiral symmetry breaking implies that
a massless (or nearly massless) quark develops a non-zero dynamical
mass (i.e. a "non-slash" term in the propagator). There are no reasons
for this quantity to be a constant independent of the momentum;
moreover, we understand that the dynamical mass should anyhow vanish at
large momenta. Its value at zero momentum can be estimated as one half
of the $\rho$ meson mass or one third of the nucleon mass, that is about

\beq
M(0)\simeq 350-400\;MeV.
\la{dynmass}\eeq
This scale is also related to chiral symmetry breaking and should emerge
together with the condensate.

The spontaneous breaking of chiral symmetry is the main dynamical
happening in QCD: it determines the face of the strong interactions
world. Indeed, it explains why the pseudoscalar mesons are light
(they are called {\em pseudo}-Goldstone bosons since they get a
non-zero mass from the small current quark masses which break somewhat
the chiral symmetry from the beginning), and why nucleons are heavy and
not degenerate in parity. The appearance of the dynamical quark mass (I
prefer this word to the constituent mass), related to the chiral
symmetry breaking, justifies, qualitatively, the old non-relativistic
or semi-relativistic models of hadrons. Indeed, once one gets
\ur{dynmass}, one can state that mesons are predominantly made of a
quark and an antiquark, and baryons are made of three quarks: an addition
of a quark-antiquark pair costs $700-800\;MeV$ which is a lot. Moreover,
the lightest vector, axial and tensor mesons and baryons should be rather
loosely bound: first, because their masses are not far away from the
sum of the dynamical quark masses, second, because, after the appearance
of a rather large scale in the theory, there is no place for the
"infrared catastrophy" --- the gluon coupling constant $\alpha_s$ remains
finite and in fact, as known from hard hadronic processes, relatively
small. Therefore, one would expect that neither the one-gluon exchange
nor the linear confining potential plays a crucial role in determining
the world of light hadrons (I shall soon cite direct evidence from
lattice experiments substantiating this view). Of course, one
would need explicit confining forces to describe highly excited resonances
lying on linear Regge trajectories, when hadrons are artificially
"stretched" by centrifugal forces, but they are not so much important
for describing low-lying hadrons. The dominant phenomenon is chiral
symmetry breaking, and if we want to understand strong interactions
we have to understand how to get \urs{chcond}{dynmass} from the
only dimensional quantity in QCD, that is $\Lambda_{QCD}$ which is the
ultraviolet cut-off-independent combination of the cut-off $\mu$ and
the gauge coupling constant $g^2(\mu)$ given at that cut-off:

\beq
\Lambda_{QCD}=\mu\exp\left[-\frac{8\pi^2}{bg^2(\mu)}\right],
\;\;\;\;b=\frac{11}{3}N_c-\frac{2}{3}N_f,
\la{Lambda}\eeq
where $N_c$ is the number of colours ($N_c=3$) and $N_f$ is the number
of active quark flavours. The value of $\Lambda_{QCD}$ depends on the
regularization scheme used; in the $\overline{MS}$ scheme
$\Lambda_{\overline{MS}}\simeq 250-300\;MeV$.

There is a remarkable puzzle, called the $U(1)$ paradox ({\sl Weinberg,
1974}):  eight out of nine pseudoscalar mesons ($\pi,\;K,\;\eta$) look
indeed like pseudo-Goldstone bosons and obey the PCAC relations while the
singlet $\eta^\prime$-meson does not (the PCAC predicts its mass to be
about $400\;MeV\;\;vs.\;\;958\;MeV\;\\(exper.)$). A related puzzle is
the $\eta\rightarrow 3\pi$ decay. It has been known from 't Hooft ({\sl
1976}) that, qualitatively, one needs {\em instantons} to cure the
$U(1)$ paradox (see also ({\sl Diakonov, 1982})). Probably the modest
$\eta^\prime$-meson tells us about strong interactions more than any
other hadron! It says that one {\em has} to take into account instanton
configurations in QCD: there is no other way to get a heavy
$\eta^\prime$-meson. We shall see that instantons bring in a most
realistic scenario of spontaneous chiral symmetry breaking as well.

\section{Instantons}

Instantons are certain configurations of the \YM potentials for the gluon
field $A_\mu^a(x)$, satisfying the equations of motion $D_\mu^{ab}
F_{\mu\nu}^b = 0$ in euclidean space, i.e. in imaginary time. The
solution has been found by Belavin, Polyakov, Schwartz and Tiupkin
({\sl 1975}); the name "instanton" has been
suggested by 't Hooft in ({\sl 1976}), who also made a
major contribution to the investigation of the instantons properties.

In QCD instantons are the best studied non-perturbative effects,
leading to the formation of the gluon condensate ({\sl Shifman,
Vainshtein and Zakharov, 1979}) and of the so-called topological
susceptibility needed to cure the $U(1)$ paradox. The QCD
instanton vacuum has been studied starting from the pioneering works in
the end of the seventies  ({\sl Callan, Dashen and Gross, 1978;
Shuryak, 1982}); a quantitative treatment of the instanton ensemble,
based on the Feynman variational principle, has been developed by
Diakonov and Petrov ({\sl 1984a}). The most striking success of the QCD
instanton vacuum is its capacity to provide a beautiful mechanism of the
spontaneous chiral symmetry breaking ({\sl Diakonov and Petrov, 1984b,
1986a}). Moreover, the instanton vacuum leads to a very reasonable
effective chiral lagrangian at low energies, including the Wess--Zumino
term, etc., which, in its turn gives a nice description of nucleons as
chiral quark solitons ({\sl Diakonov and Petrov, 1986b}).

A detailed numerical study of dozens of correlation functions in the
instanton medium undertaken by Shuryak and collaborators
({\sl 1993a}) (earlier certain correlation functions were computed
analytically by Diakonov and Petrov ({\sl 1984b, 1986b}))
demonstrated an impressing agreement with the phenomenology ({\sl
Shuryak, 1993b}) and, more recently, with direct lattice measurements
({\sl Chu et al., 1994}).  As to baryons, the instanton-motivated chiral
quark soliton model ({\sl Diakonov and Petrov, 1986b}) also leads to a
very reasonable description of dozens of baryon characteristics (see
Klaus Goeke's contribution to these Proceedings).

More recently the instanton vacuum was studied in direct lattice
experiments by the so-called cooling procedure
({\sl Teper, 1985; Ilgenfritz et al., 1986;
Polikarpov and Veselov, 1988; Chu et al., 1994}). It was
demonstrated that instantons and antiinstantons (\IIs for short)
are the only non-perturbative gluon configurations surviving
after a sufficient smearing of the quantum gluon fluctuations.  The
measured properties of the $I{\bar I}$ medium appeared ({\sl
Chu et al., 1994; Michael and Spencer, 1995}) to be close to that
computed from the variational principle ({\sl Diakonov and Petrov,
1984a}) and to what had been suggested by  Shuryak ({\sl 1982}) from
phenomenological considerations.

Cooling down the quantum fluctuations above instantons kills both the
one-gluon exchange and the linear confining potential (a small residual
string tension observed in the cooled vacuum ({\sl Polikarpov and
Veselov, 1988; Chu et al., 1994}) is probably due to the
instanton-induced rising potential at intermediate distances
({\sl Diakonov, Petrov and Pobylitsa, 1989}). Nevertheless, in the cooled
vacuum where only \IIs are left, the correlation functions of various
mesonic and baryonic currents, as well as the density-density correlation
functions representing the quark wave functions in hadrons, appear to be
quite similar to those of the true or "hot" vacuum ({\sl Chu et al.,
1994}).

In recent lectures ({\sl Diakonov, 1995}) I have reviewed the building
blocks of the instanton vacuum; an interested reader is addressed to
those lectures or original literature. Here I shall review the main facts
about instantons, in particular the way one gets chiral symmetry breaking
and the `constituent' quarks from instantons.

\section{Instanton field}

Physically, one can think of instantons in two ways: on one hand it is
a tunneling {\em process} occuring in imaginary or euclidean time (this
interpretation belongs to V.Gribov, 1976), on the other hand it is a
localized {\em pseudoparticle} in the euclidean space (A.Polyakov,
1977). The gluon field of the instanton in the singular gauge is
({\sl 't Hooft, 1976}):

\beq
A_\mu^{Ia} = \frac{2\rho^2O^{ab}\thb{i}{\mu}{\nu}(x-z)_\nu}
{(x-z)^2[\rho^2+(x-z)^2]},\;\;\;
O^{ab}=Tr(U^\dagger t^aU\sigma^i),\;\;\;O^{ai}O^{aj}=\delta^{ij}.
\la{instgena}\eeq
Here the 4-dim. vector $z_\mu$ is called the instanton centre, $\rho$
is the instanton size, the rectangular matrix $O^{ai}, \;a=1,,,(N_c^2-1),
\;i=1,2,3)$ gives the orientation of the instanton in the colour
$SU(N_c)$ space. All in all there are

\beq
4\; {\mbox (centre)}\;\;+\;\;1\; {\mbox (size)}\;\;+\;\;
(4N_c-5)\; {\mbox (orientations)}\;\;=\;\;4N_c
\la{collcoo}\eeq
so called {\em collective coordinates} desribing the field of the
instanton, of which the action is independent. The tensors
$\thb{i}{\mu}{\nu}$ are called 't Hooft symbols ({\sl 't Hooft,
1976}).

The field strength of an instanton (centered at $z_\mu=0$) is

\beq
F_{\mu\nu}^a=-\frac{4\rho^2}{(x^2+\rho^2)^2}O^{ai}\left(\thb{i}{\mu}{\nu}-
2\thb{i}{\mu}{\alpha}\frac{x_\alpha x_\nu}{x^2}-
2\thb{i}{\beta}{\nu}
\frac{x_\mu x_\beta}{x^2}\right),\;\;\;
F_{\mu\nu}^aF_{\mu\nu}^a=\frac{192\rho^4}{(x^2+\rho^2)^4},
\la{fstrength}\eeq
and satisfies the self-duality equation $F_{\mu\nu}^a=
\tilde{F}_{\mu\nu}^a$. The {\em anti-instanton} satisfies the {\em
anti}-self-dual equation, $F= -\tilde{F}$; it is given by \eqs{instgena}
{fstrength} with the replacement $\bar{\eta}\rightarrow \eta$.

The action of one (anti)instanton is

\beq
S=\frac{1}{4}\int d^4x\;F_{\mu\nu}^aF_{\mu\nu}^a=8\pi^2.
\la{act}\eeq

\section{Gluon condensate}

The QCD perturbation theory implies that the fields $A_i^a(\x)$ are
performing quantum zero-point oscillations; in the lowest order these
are just plane waves with arbitrary frequences. The aggregate energy of
these zero-point oscillations, $({\bf B}^2+{\bf E}^2)/2$, is divergent
as the fourth power of the cutoff frequency, however for any state one
has $\langle F_{\mu\nu}^2\rangle = 2\langle{\bf B}^2-{\bf E}^2\rangle =
0$, which is just a manifestation of the virial theorem for harmonic
oscillators:  the average potential energy is equal to that of the
kinetic (I am temporarily in the Minkowski space). One can prove
that this is also true in any order of the perturbation theory in the
coupling constant, provided one does not violate the Lorentz symmetry
and the renormalization properties of the theory. Meanwhile, we know
from the QCD sum rules phenomenology that the QCD vacuum posseses what
is called {\em gluon condensate} ({\sl Shifman, Vainshtein and Zakharov,
1979}):

\beq
\frac{1}{32\pi^2}\langle F_{\mu\nu}^aF_{\mu\nu}^a\rangle
\simeq (200\; MeV)^4 \;\;>\;0.
\la{glcond}\eeq

Instantons suggest an immediate explanation of this basic property of
QCD. Indeed, instanton is a tunneling process, it occurs in imaginary
time; therefore in Minkowski space one has $E_i^a=\pm iB_i^a$ (this is
actually the self-duality equation).  In euclidean space the electric
field is real as well as the magnetic one, and the gluon condensate is
just the average action density. Let us make a quick estimate of its
value.

Let the total number of \IIs in the 4-dimensional volume $V$ be $N$.
Assuming that the average separations of instantons are larger than
their average sizes (to be justified below), we can estimate the total
action of the ensemble as the sum of invidual actions:

\beq
\langle F_{\mu\nu}^2\rangle V =\int d^4x F_{\mu\nu}^2
\simeq N\cdot 32\pi^2,
\la{totact}\eeq
hence the gluon condensate is directly related to the instanton density
in the 4-dimensional euclidean space-time:

\beq
\frac{1}{32\pi^2}\langle F_{\mu\nu}^aF_{\mu\nu}^a\rangle
\simeq \frac{N}{V}\equiv \frac{1}{{\bar R}^4}.
\la{glcondest}\eeq
In order to get the phenomenological value of the condensate one needs
thus to have the average separation between pseudoparticles
({\sl Shifman, Vainshtein and Zakharov, 1979; Shuryak, 1982})

\beq
{\bar R}\simeq\frac{1}{200\;MeV}=1\;fm.
\la{avsep}\eeq

There is another point of view on the gluon condensate which I describe
briefly. In principle, all information about field theory is contained
in the partition function being the functional integral over the
fields. In the euclidean formulation it is

\beq
{\cal Z}=\int DA_\mu exp\left(-\frac{1}{4g^2}\int d^4x
F_{\mu\nu}^2\right) \stackrel{T\rightarrow\infty}{\longrightarrow}
e^{-ET},
\la{partfu}\eeq
where I have used that at large (euclidean) time $T$ the partition
function picks up the ground state or vacuum energy $E$.
For the sake of brevity I do not write the gauge fixing and
Faddeev--Popov ghost terms. If the
state is homogeneous, the energy can be written as
$E=\theta_{44}V^{(3)}$ where $\theta_{\mu\nu}$ is the stress-energy
tensor and $V^{(3)}$ is the 3-volume of the system.
Hence, at large 4-volumes $V=V^{(3)}T$ the partition function is
${\cal Z}=\exp(-\theta_{44}V)$. This $\theta_{44}$ includes zero-point
oscillations and diverges badly. A more reasonable quantity is the
partition function, normalized to the partition function understood as
a perturbative expansion about the zero-field vacuum. The
latter can be distinguished from the former by imposing a condition
that it does not contain integration over singular Yang--Mills
potentials: instanton potentials are singular at the origins. One has

\beq
\frac{{\cal Z}}{{\cal Z}_{P.T.}}= \exp\left[-(\theta_{44}
-\theta_{44}^{P.T.})V\right].
\la{pfn}\eeq

We expect that the non-perturbative vacuum energy density
$\theta_{44}-\theta_{44}^{P.T.}$ is a {\em negative} quantity since we
have allowed for tunneling: as usual in quantum mechanics, it
lowers the ground state energy. If the vacuum is isotropical, one has
$\theta_{44}=\theta_{\mu\mu}/4$. Using the trace anomaly,

\beq
\theta_{\mu\mu}\simeq
-b\frac{F_{\mu\nu}^2}{32\pi^2},\;\;\;\;
b=\frac{11}{3}N_c,
\la{TA}\eeq
one gets ({\sl Diakonov and Petrov, 1984a}):

\beq
\frac{{\cal Z}}{{\cal Z}_{P.T.}}=\exp\left( \frac{b}{4}V
\langle F_{\mu\nu}^2/32\pi^2\rangle_{NP}\right)
\la{GCdef}\eeq
where $\langle F_{\mu\nu}^2\rangle_{NP}$ is the gluon field vacuum
expectation value which is due to non-perturbative fluctuations, i.e.
the gluon condensate. The aim of any QCD-vacuum builder
is to minimize the vacuum energy or, equivalently, to maximize the
gluon condensate.

\section{One-instanton weight}

The words `instanton vacuum' mean that one assumes that the QCD
partition function is mainly saturated by an ensemble of interacting \IIs
together with quantum fluctuations about them. Instantons are
necessarily present in the QCD vacuum if only because they lower the
vacuum energy in respect to the purely perturbative (divergent) one.
The question is whether they give the dominant contribution to the
gluon condensate, and to other basic quantities. To answer this
question one has to compute the partition function \ur{partfu} assuming
that it is mainly saturated by instantons, and to compare the
obtained gluon condensate with the phenomenological one. This work
has been done a decade ago in ref. ({\sl Diakonov and Petrov, 1984a});
today direct lattice measurements seem to confirm that the answer to the
question is positive: the observed density of \IIs is in agreement with
the estimate \ur{avsep}.

The starting point of this calculation is the contribution of one
isolated instanton to the partition function \ur{partfu}, or the
one-instanton weight. To get a reasonable result it must be {\it i})
normalized (to the determinant of the free quadratic form, i.e. with no
background field), {\it ii}) regularized (for example by using the
Pauli--Villars method), and {\it iii}) accounted for the zero modes.
The resulting one-instanton contribution to the partition function
(normalized to the free one) is an integral over the $4N_c$ collective
coordinates of an instanton ({\sl 't Hooft, 1976; Bernard, 1979}):

\beq
\frac{{\cal Z}_{1-inst}}{{\cal Z}_{P.T.}}
=\int d^4z_\mu\int d\rho\int dO\;
\frac{C(N_c)}{\rho^5}\left[\frac{8\pi^2}{g^2(\mu)}\right]^{2N_c}
(\mu\rho)^{\frac{11}{3}N_c}\exp\left(-\frac{8\pi^2}{g^2(\mu)}\right)
\la{cuto}\eeq
\beq
=\int d^4z_\mu\int d\rho\int dO\;
\frac{C(N_c)}{\rho^5}\left[\frac{8\pi^2}{g^2(\mu)}\right]^{2N_c}
(\rho\Lambda_{QCD})^{\frac{11}{3}N_c}.
\la{1instw}\eeq

The product of the last two factors in \eq{cuto} is actually a
combination of the cut-off $\mu$ and the bare coupling constant $g^2(\mu)$
given at this cut-off, which is cutoff-independent; it can be replaced
by $(\rho\Lambda_{QCD})^{11N_c/3}$, see \eq{Lambda}. This is the way
$\Lambda_{QCD}$ enters into the game; henceforth all dimensional
quantities will be expressed through $\Lambda_{QCD}$, which is, of
course, a welcome message. The numerical coefficient $C(N_c)$ depends
explicitly on the number of colours; it also implicitly depends on the
regularization scheme used.

Note that the $g^2$ in the pre-exponent starts to "run" only
at the 2-loop level, hence its argument is taken at the ultra-violet
cut-off $\mu$. The 2-loop instanton weight can be found in refs.
({\sl Diakonov, 1995; Diakonov, Polyakov and Weiss, 1955}).

In both one- and two-loop approximations the integral over the  instanton
sizes $\rho$ in \eq{1instw} diverges as a high power of $\rho$ at large
$\rho$: this is of course the consequence of asymptotic freedom. It means
that individual instantons tend to swell. This circumstance plagued the
instanton calculus for many years. If one attemts to cut the $\rho$
integrals "by hand", one violates the renormalization properties of the
YM theory. Actually the size integrals appear to be cut from above due to
instanton interactions.

\section{Instanton ensemble}

To get a volume effect from instantons one needs to consider an
\II ensemble, with their total number $N$ proportional to the
4-dimensional volume $V$. Immediately a mathematical difficulty arises:
any superposition of \IIs is not, strictly speaking, a solution of the
equation of motion, therefore, one cannot directly use the
semiclassical approach of the previous section. There are two ways to
overcome this difficulty. One is to use a variational principle
({\sl Diakonov and Petrov, 1984a}), the other is to use the effective YM
lagrangian in the instanton field ({\sl Diakonov and Polyakov, 1993}).

The idea of the variational principle is to use a modified YM action
for which a chosen \II ansatz {\em is} a saddle point. Exploiting the
convexity of the exponent one can prove that the true vacuum energy is
{\em less} than that obtained from the modified action. One can
therefore use variational parameters (or even functions) to get a best
upper bound for the vacuum energy. We call it the Feynman variational
principle since the method was suggested by Feynman in his famous
study of the polaron problem. Todays direct lattice investigation of the
\II ensemble seem to indicate that Petrov and I have obtained rather
accurate numbers in this terrible problem.  Therefore I will cite the
numerics from those calculations in what follows.

The main finding is that the \II ensemble stabilizes at a certain
density related to the $\Lambda_{QCD}$ parameter (there is no other
dimensional quantity in the theory!):

\beq
\langle F_{\mu\nu}^2/32\pi^2\rangle
\simeq \frac{N}{V}\geq (0.75 \Lambda_{\overline{MS}})^4
\la{numval}\eeq
which would require $\Lambda_{\overline{MS}}\simeq 270\;MeV$ to get
the phenomenological value of the condensate. It should be mentioned
however that using more sophisticated variational Ans\"{a}tze one can
obtain a larger coefficient in \eq{numval} and hence would need smaller
values of $\Lambda$.

The average sizes $\bar \rho$ appears to be much less
than the average separation $\bar R$.  Numerically we have found for
the $SU(3)$ colour:

\beq
\frac{\bar \rho}{\bar R} \simeq \frac{1}{3}
\la{pf}\eeq
which coincides with what was suggested previously by Shuryak
({\sl 1982}) from considering the phenomenological applications of
the instanton vacuum. This value should be compared with that found from
direct lattice measurements ({\sl Chu et al., 1994}): $\bar \rho /\bar R
\simeq .37 - .4$, depending on where one stops the cooling procedure. The
packing fraction, i.e. the fraction of the 4-dimensional volume
occupied by instantons apears thus to be rather small, $\pi^2 \bar\rho^4
/\bar R^4 \simeq 1/8$. This small number can be traced back to the
"accidentally" large numbers appearing in the 4-dimensional YM theory:
the $11N_c/3$ of the Gell-Mann--Low beta function and the number of
zero modes being $4N_c$. Meanwhile, it is exactly this small packing
fraction of the instanton vacuum which gives an {\it a posteriori}
justification for the use of the semi-classical methods. As I shall show
in the next sections, it also enables one to identify adequate degrees of
freedom to describe the low-energy QCD.

\section{Chiral symmetry breaking by instantons: qualitative
deri\-va\-tion}

The key observation is that the Dirac operator in the background field
of one (anti) instanton has an exact zero mode with $\lambda=0$
({\sl 't Hooft, 1976}). It is a consequence of the general Atiah--Singer
index theorem; in our case it is guaranteed by the unit Pontryagin
index or the topological charge of the instanton field. These
zero modes are 2-component Weyl spinors: {\em right}-handed for
instantons and {\em left}-handed for antiinstantons.

For infinitely separated $I$ and $\bar I$ one has thus
two degenerate states with exactly zero eigenvalues. As usual in
quantum mechanics, this degeneracy is lifted through the
diagonalization of the hamiltonian, in this case the hamiltonian is
the full Dirac operator. The two "wave functions" which diagonalize the
"hamiltonian" are the sum and the difference of the would-be zero
modes, one of which is a 2-component left-handed spinor $\Phi_1$, and
the other is a 2-component right-handed spinor $\Phi_2$. The resulting
wave functions are 4-component Dirac spinors; one can be obtained from
another by multiplying by the $\gamma_5$ matrix.  As the result the two
would-be zero eigenstates are split symmetrically into two
$4$-component Dirac states with {\em non-zero} eigenvalues equal to the
overlap integral between the original states $\Phi_{1,2}$:

\beq
\lambda=\pm |T_{12}|,\;\;\;\;
T_{12}=\int d^4x \Phi_1^\dagger(-i\!\not\partial)\Phi_2
\stackrel{R_{12}\rightarrow\infty}{\longrightarrow}
-\frac{2\rho_1\rho_2}{R_{12}^4}{\mbox Tr}(U_1^\dagger U_2R_{12}^+).
\la{overl}\eeq

We see that the splitting between the would-be zero modes fall off as
the third power of the distance between $I$ and $\bar I$; it also
depends on their relative orientation.

When one adds more \IIs each of them brings in a would-be zero
mode. After the diagonalization they get split symmetrically in
respect to the $\lambda=0$ axis.  Eventually, for an \II ensemble one
gets a continuous band spectrum with a spectral density $\nu(\lambda)$
which is even in $\lambda$ and finite at $\lambda=0$. Meanwhile,
there is a general formula relating the chiral condensate with the
average spectral density of the Dirac operator at zero $\lambda$ ({\sl
Banks and Casher, 1980; Diakonov and Petrov, 1984b}):

\beq
\langle \bar\psi \psi\rangle = -\frac{1}{V}\mbox{sign}(m)\;\pi\;
\overline{\nu(0)}
\la{BC}\eeq
Since \IIs lead to a non-zero $\nu(0)$ they automatically give the
non-zero chiral condensate, which signals chiral symmetry breaking.

One can make a quick estimate of $\langle\bar\psi\psi\rangle$: Let the
total number of \IIs in the 4-dimensional volume $V$ be $N$. The spread
$\Delta$ of the band spectrum of the would-be zero modes is given by
their average overlap \ur{overl}:

\beq
\Delta\sim\sqrt{\int (d^4R/V) T(R)T^*(R)}\sim\frac{\bar\rho}{\bar R^2}
\la{averoverl}\eeq
where $\bar\rho$ is the average size and $\bar R=(N/V)^{-1/4}$ is the
average separation of instantons. Note that the spread of the
would-be zero modes is parametrically much less than $1/\bar\rho$ which
is the typical scale for the non-zero modes. Therefore, neglecting the
influence of the non-zero modes is justified if the packing
fraction of instantons is small enough. From \eq{BC} one gets an
estimate for the chiral condensate induced by instantons:

\beq
\langle\bar\psi\psi\rangle=-\frac{\pi}{V}\nu(0)
\simeq -\frac{\pi}{V}\frac{N}{\Delta}\sim-\frac{1}{\bar R^2 \bar\rho}.
\la{ccest}\eeq

It is amusing that the physics of the spontaneous breaking of chiral
symmetry resembles the so-called Anderson conductivity in
disordered systems. Imagine random impurities (atoms) spread over a
sample with final density, such that each atom has a localized bound
state for an electron. Due to the overlap of those localized electron
states belonging to individual atoms, the levels are split into a band,
and the electrons become delocalized. That means conductivity of the
sample. In our case the localized zero quark modes of individual
instantons randomly spread over the volume get delocalized due to their
overlap, which means chiral symmetry breaking.

I should mention that the idea that instantons can break chiral
symmetry has been discussed previously (see  refs. ({\sl
Caldi, 1977; Carlitz and Creamer, 1978; Callan, Dashen and Gross,
1978, Shuryak, 1982}), however the present mechanism and a consistent
formalism has been suggested and developed in the papers ({\sl Diakonov
and Petrov, 1984b, 1986a}).

\section{Chiral symmetry breaking by instantons: quark propa\-gator}

Having explained the physical mechanism of chiral symmetry breaking as
due to the delocalization of the would-be zero fermion modes in the
field of individual instantons, I shall indicate how to compute
observables in the instanton vacuum. The main quantity is the quark
propagator in the instanton vacuum, averaged over the instanton
ensemble. This quantity has been calculated in ({\sl Diakonov and
Petrov, 1984b; Pobylitsa, 1989}). In particular, Pobylitsa has
derived a closed equation for the averaged quark propagator, which can be
solved as a series expansion in the formal parameter $N\bar\rho^4/VN_c$
which numerically is something like 1/250.

The result of these works is that in the leading order in the  above
parameter the quark propagator has the form of a massive propagator with
a momentum-dependent dynamical mass:

\beq
S(p)=\frac{\not p +iM(p^2)}{p^2+M^2(p^2)}, \;\;\;M(p^2)
=c\sqrt{\frac{\pi^2N\bar\rho^2}{VN_c}}F(p\bar\rho),
\la{propag}\eeq
where $F(z)$ is a combination of the modified Bessel functions and
is related to the Fourier transform of the quark zero mode: it is
equal to one at $z=0$ and decreases rapidly with the momentum, measured
in units of the inverse average size of instantons;  $c$ is a constant of
the order of unity which depends slightly on the approximation used in
deriving the propagator. Note that the dynamical quark mass is
non-analytical in the instanton density.

Fixing the average  density by the empirical gluon condensate (see
section 4) so that $\bar R\simeq 1\;fm$ and fixing the ratio
$\bar\rho/\bar R = 1/3$ from our variational estimate, we get the value
of the dynamical mass at zero momentum,

\beq
M(0) \simeq (350)\;MeV
\la{massatz}\eeq
while the quark condensate is

\beq
\langle\bar\psi\psi\rangle=i\int\frac{d^4p}{(2\pi)^4}{\rm Tr} S(p)
=-4N_c\int\frac{d^4p}{(2\pi)^4}\frac{M(p)}{p^2+M^2(p)}
\simeq - (255\;MeV)^3.
\la{ccn}\eeq
Both numbers, \ur{massatz} and \ur{ccn}, appear to be close to their
phenomenological values.

Using the above small parameter one can also compute more complicated
quantities like 2- or 3-point mesonic correlation functions of the type

\beq
\langle J_A(x)J_B(y)\rangle, \;\;\;\langle J_A(x)J_B(y)J_c(z)\rangle,
\;\;\;J_A=\bar\psi\Gamma_A\psi
\la{mescorr}\eeq
where $\Gamma_A$ is a unit matrix in colour but an arbitrary matrix in
flavour and spin. Instantons influence the correlation function in
two ways: {\it i}) the quark and antiquark propagators get dressed and
obtain the dynamical mass, as in \eq{propag}, {\it ii}) quark and
antiquark may scatter simultaneously on the same pseudoparticle;
that leads to certain effective quark interactions. These interactions
are strongly dependent on the quark-antiquark quantum numbers: they are
strong and attractive in the scalar and especially in the pseudoscalar
and the axial channels, and rather weak in the vector and tensor
channels.

Since we have already obtained chiral symmetry breaking by studying a
single quark propagator in the instanton vacuum, we are doomed to have
a massless Goldstone pion in the psuedoscalar and axial correlators.
Having a concrete dynamical realization of chiral symmetry
breaking at hand, we can not only check the general Ward identities  of
the PCAC (which work of course) but we are in a position to find
quantities whose values do not follow from general relations. One of the
most important quantities is the $F_\pi$ constant: it can be calculated
as the residue of the pion pole. We get:

\beq
F_\pi=\frac{{\rm const}}{\bar\rho}\left(\frac{\bar\rho}{\bar
R}\right)^2\sqrt{\ln\frac{\bar R}{\bar\rho}}\simeq 100\;MeV\;\;\;{\rm
vs.}\;\; 93\;MeV\;\;({\rm exper.}).
\la{Fpi}\eeq
This is a very instructive formula. The point is, $F_\pi$ is
surprisingly small in the strong interactions scale which, in the
instanton vacuum, is given by the average size of pseudoparticles,
$1/\bar\rho \simeq 600\;MeV$. The above formula says that $F_\pi$ is
down by the packing fraction factor $(\bar\rho/\bar R)^2\simeq 1/9$.
It can be said that $F_\pi$ measures the diluteness of the instanton
vacuum! However it would be wrong to say that instantons are in a
dilute gas phase -- the interactions are crucial to stabilize the
medium and to support the known renormalization properties of the
theory, therefore they are rather in a liquid phase, however dilute it
may turn to be.

By calculating three-point correlation functions in the instanton
vacuum we are able to determine, e.g. the charge radius of the
Goldstone excitation:

\beq
\sqrt{r_\pi^2}\simeq \frac{\surd N_c}{2\pi F_\pi}\simeq (340\;MeV)^{-1}
\;\;\;{\rm vs.}\;\;(310\;MeV)^{-1}\;\;({\rm exper.}).
\la{chra}\eeq

A systematic numerical study of various correlation functions in the
instanton vacuum has been performed by Shuryak and collaborators
({\sl 1993a}). In all cases considered the results
agree well or very well with experiments and phenomenology. As I already
mentioned in the introduction, similar conclusions have been recently
obtained from direct lattice measurements ({Chu et al., 1994}).

\section{Chiral symmetry breaking by instantons: Nambu--Jona-Lasinio
model}

The idea of the first two derivations of chiral symmetry breaking by
instantons, presented above, is: "Calculate quark observables in a
given  background gluon field, then average over the ensemble of
fields", in our case the ensemble of \IIs. The idea of the third
derivation is the opposite: "First average over the \II ensemble
and obtain an effective theory written in terms of interacting quarks
only. Then compute observables from this effective theory". This
approach is in a sense more economical; it has been developed in
({\sl Diakonov and Petrov, 1986a; Diakonov, 1995; Diakonov, Polyakov and
Weiss, 1995}).

Quark interaction arises when two or more quarks happen to
scatter on the same pseudoparticle; averaging over its positions and
orientations results in a four- (or more) fermion interaction term
whose range is that of the average size of instantons. The most
essential way how instantons influence quarks is, of course, via the
zero modes. Since each massless quark flavour has its own zero mode,
it means that the effective quark interactions will be actually
$2N_f$ fermion ones. They are usually referred to as {\em 't Hooft
interactions} as he was the first to point out the quantum numbers of
these effective instanton-induced interactions ({\sl 't Hooft, 1976}).
In case of two flavours they are four-fermion interactions, and the
resulting low-energy theory resembles the old
Vaks--Larkin--Nambu--Jona-Lasinio model ({\sl Vaks et al., 1961}) which
is known to lead to chiral symmetry breaking.

Let me quote the result for the case of two flavours, $N_f=2$
({\sl Diakonov and Petrov, 1986a}). In that case one gets a 4-quark
interaction vertex:

\[
Y_2^{(+)}=\frac{2N_c^2}{N/V}\int\frac{d^4k_1d^4k_2d^4l_1d^4l_2}{(2\pi)^{12}}
\sqrt{M(k_1)M(k_2)M(l_1)M(l_2)}
\]
\[
\cdot\frac{\epsilon^{f_1f_2}\epsilon_{g_1g_2}}{2(N_c^2-1)}\left[
\frac{2N_c-1}{2N_c}(\psi_{Lf_1}^\dagger(k_1)\psi_L^{g_1}(l_1))
(\psi_{Lf_2}^\dagger(k_2)\psi_L^{g_2}(l_2))\right.
\]
\beq
\left. +\frac{1}{8N_c}
(\psi_{Lf_1}^\dagger(k_1)\sigma_{\mu\nu}\psi_L^{g_1}(l_1))
(\psi_{Lf_2}^\dagger(k_2)\sigma_{\mu\nu}\psi_L^{g_2}(l_2))\right]
\la{Y2}\eeq
where $\psi_L^f$ is the left-handed component of the quark of flavour
$f\;(f=u,d)$. For the antiinstanton-induced vertex $Y^{(-)}$ one has to
replace left handed components by right-handed. $M(k)$ is the same
dynamical quark mass as obtained in another approach outlined in the
previous section. It is given by ($z=k\rho/2$):

\[
M(k)=M(0)F^2(k),\;\;\;M(0)=\lambda (2\pi\rho)^2,
\]
\beq
F(k)=2z[I_0(z)K_1(z)-I_1(z)K_0(z)-\frac{1}{z}I_1(z)K_1(z)
\stackrel{k \rightarrow\infty}{\longrightarrow} \frac{6}{k^3\rho^3},
\;\;\;\;F(0)=1.
\la{dmass}\eeq

The value $M(0)$ (or $\lambda$) is found from the equation
({\sl Diakonov and Petrov, 1984b, 1986a}) (called sometimes
self-consistency or gap equation):

\beq
\frac{4N_c}{N/V}\int\frac{d^4k}{(2\pi)^4}\frac{M^2(k)}{k^2+M^2(k)}=1.
\la{gap}\eeq
It is seen from \eq{gap} that the dynamically
generated mass is of the order of $M(0)\sim \sqrt{N/(VN_c)}\bar\rho$.
Knowing the form of $M(k)$ given by \eq{dmass} and using the "standard"
values $N/V=(1\;{\rm fm})^{-4},\; \bar\rho=(1/3)$ fm we find
numerically $M(0)\simeq 350$ MeV. If one neglects $M^2$ in the
denominator of \eq{gap} ({\sl Pobylitsa, 1989}) one gets $M(0)\simeq 420$
MeV.  This deviation indicates the accuracy of the "zero mode
approximation" used in this derivation: it is about 15\%.

Note that the second (tensor) term in \eq{Y2} is negligible at large
$N_c$.  Using the identity

\beq
2\epsilon^{f_1f_2}\epsilon_{g_1g_2}=\delta_{g_1}^{f_1}\delta_{g_2}^{f_2}
-(\tau^A)_{g_1}^{f_1}(\tau^A)_{g_2}^{f_2}
\la{id}\eeq
one can rewrite the leading (first) term of \eq{Y2} as

\beq
(\psi^\dagger\psi)^2+(\psi^\dagger\gamma_5\psi)^2-(\psi^\dagger\tau^A\psi)^2
-(\psi^\dagger\tau^A\gamma_5\psi)^2
\la{NJLf}\eeq
which resembles closely the Nambu--Jona-Lasinio model. It should be
stressed though that in contrast to that {\em at hoc} model the
interaction \ur{Y2} {\it i}) violates explicitly the $U_A(1)$ symmetry,
{\it ii}) has a fixed interaction strength and {\it iii}) contains an
intrinsic ultraviolet cutoff due to the formfactor function $M(k)$.
This model is known to lead to chiral symmetry breaking, at least at
large $N_c$ when the use of the mean field approximation to the model
is theoretically justified.

The bosonization of these interactions has been performed in
({\sl Diakonov and Petrov, 1986a}); it paves the way to
studying analytically various correlation functions in the instanton
vacuum.

A separate issue is the application of these ideas to hadrons made
of light and heavy quarks ({\sl Chernyshev, Nowak and Zahed, 1994}).

\section{QCD at still lower energies}

Using the packing fraction of instantons $\bar\rho/\bar R\simeq 1/3$ as
a new algebraic parameter one observes that all degrees of freedom in
QCD can be divided into two categories: {\it i}) those with masses $\ge
1/\bar\rho$ and {\it ii}) those with masses $\ll 1/\bar\rho$. If one
restricts oneself to low-energy strong interactions such that
momenta are $\ll 1/\bar\rho\simeq 600\;MeV$, one can neglect the
former and concentrate on the latter. There are just two kind of
degrees of freedom whose mass is much less than the inverse average
size of instantons: the (pseudo) Goldstone pseudoscalar
mesons and the quarks themselves which obtain a dynamically-generated
mass $M\sim (1/\bar\rho)(\bar\rho^2/\bar R^2)\ll 1/\bar\rho$. Thus in
the domain of momenta $k\ll 1/\bar\rho\;$ QCD reduces to a remarkably
simple though nontrivial theory of massive quarks interacting with
massless or nearly massless pions. It is given by the partition
function ({\sl Diakonov and Petrov, 1984b, 1986a})

\beq
{\cal Z}_{QCD}^{{\rm low\; mom.}}=\int D\psi D\psi^\dagger
\exp\left[\int d^4x
\psi^\dagger\left(i\!\not\partial+iMe^{i\pi^A\tau^A\gamma_5}\right)
\psi\right].
\la{lowmom}\eeq
Notice that there is no kinetic energy term for the pions, and that
the theory is not a renormalizable one. The last circumstance is due to
the fact that it is an effective low-energy theory; the ultraviolet
cutoff is actually $1/\bar\rho$.

There is a close analogy with solid state physics here. The microscopic
theory of solid states is QED: it manages to break spontaneously the
translational symmetry, so that a Goldstone excitation emerges, called the
phonon. Electrons obtain a "dynamical mass" $m^*$ due to hopping from
one atom in a lattice to another. The "low energy" limit of solid
state physics is described by interactions of dressed electrons
with Goldstone phonons.  These interactions are more or less fixed by
symmetry considerations apart from a few constants which can be deduced
from experiments or calculated approximately from the underlying QED.
Little is left of the complicated dynamics at the atom scale.

What Petrov and I have attempted, is a similar path: one starts with
the fundamental QCD, finds that instantons stabilize at a relatively low
density and that they break chiral symmetry; what is left at low
momenta are just the dynamically massive quarks and massless pions.
One needs two scales to describe strong interactions at low momenta:
the ultra-violet cutoff, whose role is played by the inverse instanton
size, and the dynamical quark mass proportional to the square
of the instanton density. If one does not believe our variational
calculations of these quantities one can take them from experiment.

If one integrates off the quark fields in \eq{lowmom} one gets the
{\em effective chiral lagrangian}:

\[
S_{eff}[\pi]=-N_c{\rm Tr\;ln}\left(i\!\not\partial+iMU^{\gamma_5}\right),
\]
\beq
U=\exp(i\pi^A\tau^A),\;\;\;U^{\gamma_5}=\exp(i\pi^A\tau^A\gamma_5),
\;\;\;L_\mu=iU^\dagger\partial_\mu U.
\la{chilagr}\eeq

One can expand \eq{chilagr} in powers of the derivatives of the pion
field and get:

\[
S_{eff}[\pi]=\frac{F_\pi^2}{4}\int d^4x\;{\rm Tr}L_\mu^2
-\frac{N_c}{192\pi^2}\int d^4x \left[2{\rm Tr}(\partial_\mu L_\mu)^2
+{\rm Tr}L_\mu L_\nu L_\mu L_\nu\right]
\]
\beq
+\frac{N_c}{240\pi^2}\int
d^5x\;\epsilon_{\alpha\beta\gamma\delta\epsilon}\;
{\rm Tr}L_\alpha L_\beta L_\gamma L_\delta L_\epsilon +...
\la{derexpan}\eeq
The first term here is the old Weinberg chiral lagrangian with

\beq
F_\pi^2=4N_c\int\frac{d^4k}{(2\pi)^4}\frac{M^2(k)}{[k^2+M^2(k)]^2};
\la{Fpieq}\eeq
the second term are the four-derivative Gasser--Leutwyler terms
(with coefficients which turn out to agree with those following from
the analysis of the data); the last term in \eq{derexpan} is the
so-called Wess--Zumino term.  Note that the $F_\pi$ constant diverges
logarithmically at large momenta; the integral is cut by the momentum-
dependent mass at $k\sim 1/\bar\rho$, so that one gets the same
expression as in a different approach described in section 10, see
\eq{Fpi}.

An ideal field of application of the low-momentum partition function
\ur{lowmom} is the quark-soliton model of nucleons ({\sl Diakonov and
Petrov, 1986b}) -- actually the model has been derived from this
partition function. The size of the nucleon $\sim (250\; MeV)^{-1}$ is
much larger than the size of instantons $\sim (600\; MeV)^{-1}$; hence
the low-momentum theory \ur{lowmom} seems to be justified. Indeed, the
computed static characteristics of baryons like formfactors, magnetic
moments, etc., are in a good accordance with the data (see the talk
by K.Goeke).

\section{Can instantons give confinement?}

Our analytical calculations sketched in these lectures, the extensive
numerical studies of the instanton vacuum by Shuryak and collaborators
and the recent direct lattice measurements -- all point out that
instantons play a crucial role in determining the world of light
hadrons, including the nucleon. Confinement has not much to do with it
-- contrary to what has been a common wisdom a decade ago and in what
many people still believe. Nevertheless, confinement is a property of
QCD, and one needs to understand the confinement mechanism. What can
be said today is that confinement must be "soft": it should destroy
neither the successes of the perturbative description of high-energy
processes (no "string effects" there) nor the successes of instantons
at low momenta.

Quite recently we have noticed ({\sl Diakonov and Petrov, 1995}) that
one can obtain the area behaviour of large Wilson loops (i.e.
confinement) from instantons, provided their size distribution behaves
as $\sim d\rho/\rho^3$ for large $\rho$. It is remarkable that only
the large-distance behaviour of the interquark potential is sensitive to
the tail of the size distribution -- all other quantities discussed
in this paper are determined rather by the average sizes. Moreover,
precisely the "one over cube" distribution law seems to follow from
the statistical mechanics of instantons, when one takes into account
the quantum gluon fluctuations in the instanton ensemble.

My special gratitude is to Victor Petrov with whom we worked together for
many years on the topics discussed in this talk.

\vskip .5true cm

\hskip 1.5true cm {\bf References}\\
\vskip .2true cm
Banks, T. and A.Casher (1980). {\it Nucl. Phys.} {\bf B169}, 103
\\
Belavin, A., A.Polyakov, A.Schwartz and Yu.Tyupkin (1975). Pseudoparticle
solutions of the Yang--
\parindent .4cm
\parskip 0cm

Mills equations. {\it Phys. Lett.}
{\bf 59}, 85; A.Polyakov. Quark confinement and topology of gauge

fields.  {\it Nucl. Phys.} {\bf B120} (1977) 429.  \\
Bernard, C. (1979). {\it Phys. Rev.} {\bf D19}, 3013. \\
Caldi, D.G. (1977). {\it Phys. Rev.  Lett.} {\bf 39}, 121. \\
Callan, C., R.Dashen and D.Gross (1978). Toward a theory of the strong
interactions. {\it Phys. Rev.} {\bf D17}, 2717. \\
Carlitz, R.D. (1978). {\it Phys. Rev.} {\bf D17}, 3225; R.D.Carlitz
and D.B.Creamer , {\it Ann. Phys. (N.Y.)}

{\bf 118} (1979) 429. \\
Chernyshev, S., M.Nowak and I.Zahed (1994). Heavy mesons in a random
instanton gas. Preprint

SUNY-NTG-94-37. \\
Chu, M.-C.,J.Grandy, S.Huang and J.Negele (1994).  {\it Phys. Rev.}
{\bf D49}, 6039; {\it Phys. Rev. Lett.}

{\bf 70} (1993) 225. \\
Diakonov, D. (1983). The U(1) problem and instantons.  In: {\it Gauge
Theories of the Eighties}, Lecture

Notes in Physics, Springer-Verlag, p.207. \\
Diakonov, D., and V.Petrov (1984a).  Instanton-based vacuum
from the Feynman variational

principle. {\it Nucl. Phys.} {\bf B245}, 259. \\
Diakonov, D. and V.Petrov (1984b).  Chiral condensate in the
instanton vacuum. {\it Phys. Lett.} {\bf 147B}

351; {\it Sov. Phys.
JETP} {\bf 62} (1985) 204, 431; A theory of light quarks in the
instanton vacuum.

{\it Nucl.Phys.} {\bf B272} (1986) 457. \\
Diakonov, D. and V.Petrov (1986a). Spontaneous breaking of chiral symmetry in the
instanton

vacuum. Preprint LNPI-1153 (1986), published (in Russian) in:
{\it Hadron matter under extreme

conditions}, Kiew (1986) p.192;
Diquarks in the instanton picture.  In: {\it Quark Cluster Dynamics},

Lecture Notes in Physics, Springer-Verlag (1992) p.288. \\
Diakonov, D. and V.Petrov (1986b).  Chiral theory of nucleons.  {\it
Sov.  Phys.  JETP.  Lett.} {\bf 43}, 75;

Diakonov, D., V.Petrov and
P.Pobylitsa. A chiral theory of nucleons {\it Nucl.  Phys.} {\bf B306}
(1988) 809;

D.Diakonov, V.Petrov and M.Praszalowicz. Nucleon mass
and nucleon sigma term. {\it Nucl.  Phys.}

{\bf B323} (1989) 53. \\
Diakonov, D., V.Petrov and P.Pobylitsa (1989). The Wilson loop and
heavy-quark potential in the

instanton vacuum. {\it Phys.  Lett.} {\bf
226B}, 372.  \\
Diakonov, D.  and M.Polyakov (1993). Baryon number non-conservation  at
high energies and instan\-

ton interactions. {\it Nucl. Phys.} {\bf B389}, 109. \\
Diakonov, D., M.Polyakov and C.Weiss (1995).  Hadronic matrix elements
of  gluon operators in the

instanton vacuum. hep-ph/9510232,
to be published in {\it Nucl. Phys.  B}. \\
Diakonov, D. (1995). Chiral symmetry breaking by instantons.
{\it Lectures at the Enrico Fermi School},

Varenna, June 27 -- July 7, 1995. hep-ph/9602375, to be published
in the Proceedings \\
Diakonov, D. and V.Petrov (1995).  Confinement from instantons? In:
{\it Proceedings of the Interna-

tional workshop}, Trento, July 1995, to be published. \\
't Hooft, G. (1976).  Symmetry breaking through Adler--Bell--Jackiw
anomalies. {\it Phys.  Rev. Lett.}

{\bf 37}, 8; Computation of quantum
effects due to a four dimensional pseudoparticle. {\it Phys. Rev.} {\bf
D14}

(1976) 3432. \\
Ilgenfritz, E.-M., M.Laursen, M.M\"{u}ller-Preussker, G.Shierholz and
H.Schil\-ler (1986).  {\it Nucl.  Phys.}

{\bf B268}, 693. \\
Michael, C. and P.Spencer (1995).  Cooling in the $SU(2)$ instanton vacuum.
Helsinki preprint HU

TFT 95-21, Liverpool preprint LTH-346 (1995), hep-lat/9503018. \\
Pobylitsa, P. (1989). The quark propagator and
correlation functions in the instanton vacuum. {\it Phys.

Lett.} {\bf 226B} 387. \\
Polikarpov, M. and A.Veselov (1988). {\it Nucl. Phys.} {\bf B297}, 34.
\\ Shifman, M., A.Vainshtein and V.Zakharov (1979). QCD and resonance
physics.  Theoretical foun\-

dations. {\it Nucl.Phys.} {\bf B147}, 385. \\ Shuryak, E. (1982). The
role of instantons in quantum chromodynamics, I, II, III. {\it Nucl.
Phys.} {\bf B203}

93, 116, 140.  \\
Shuryak, E.  (1993a). E.Shuryak and J.Verbaarschot {\it Nucl.  Phys.}
{\bf B410} (1993) 55;
T.Sch\"{a}fer,

E.Shuryak and J.Verbaarschot, {\it Nucl.  Phys.} {\bf
B412} (1994) 143; T.Sch\"{a}fer and E.Shuryak, {\it Phys.

Rev.} {\bf D50} (1994) 478. \\
Shuryak, E.  (1993b). {\it Rev. Mod.  Phys.} {\bf 65}, 1. \\
Teper, M.  (1985). {\it Phys. Lett.} {\bf 162B}, 357. \\
Vaks, V.G. and A.I.Larkin (1961). {\it ZhETF} {\bf 40}, 282; Y.Nambu
and G.Jona-Lasinio. {\it Phys. Rev.} {\bf 122}

(1961) 345. \\
Weinberg, S. (1974). $U(1)$ problem. In: {\it Proc. XVII Internat. Conf.
on High Energy Physics},

London, v.3 p.59.

\end{document}